# The COVID-19 vaccination, preventive behaviors and pro-social motivation: panel data analysis from Japan

Short title: COVID-19 vaccination and preventive behaviors

Eiji Yamamura[1*], Yoshiro Tsutsui[2], Fumio Ohtake[3]

[1] Department of Economics, Seinan Gakuin University, Fukuoka, Japan

*Corresponding author
Email: yamaei@seinan-gu.ac.jp (EY)
Full list of author information is available at the end of the article

# Abstract

## Background

The COVID-19 vaccine reduces infection risk: even if one contracts COVID-19, the probability of complications like death or hospitalization is lower. However, vaccination may prompt people to decrease preventive behaviors, such as staying indoors, handwashing, and wearing a mask. Thereby, if vaccinated people pursue only their self-interest, the vaccine's effect may be lower than expected. However, if vaccinated people are pro-social (motivated toward benefit for the whole society), they might maintain preventive behaviors to reduce the spread of infection.

## Methods

We conducted 26 surveys nearly once a month from March 2020 (the early stage of COVID-19) to September 2022 in Japan. By corresponding with the identical individuals, we independently constructed original panel data (N = 70,979). Based on the data, we identified the timing of the second vaccine shot and compared preventive behaviors before and after vaccination. We investigated whether second-shot vaccination correlated with changes in preventive behaviors. Furthermore, we explored whether the vaccination effect differs between senior and younger groups. We then investigated the effect of pro-social motivation on preventive behaviors.

## Results

Major findings are as follows: (1) Being vaccinated led people to increase preventive behaviors, such as mask-wearing by 1.04 (95% confidence intervals [Cis]: 0.96–1.11) points, and handwashing by 0.34 (95% CIs: 0.30–0.38) points, on a 5-point scale. (2) vaccinated people under the age of 65 years are less likely to stay indoors. (3) people with pro-social motivation to be vaccinated are more likely to maintain prevention than those not so motivated; the difference is 0.08 (95% CIs: 0.01–0.15) points for mask-wearing and 0.05 (95% CIs: 0.001–0.10) points for handwashing, on a 5-point scale.

## Conclusion

After vaccination, the opportunity cost of staying indoors outweighs its benefits and people are less inclined to stay home. This effect is lower for older people, who are at higher risk of serious illness. The opportunity cost of mask-wearing and handwashing is lower than that of staying indoors, and the benefit persists after vaccination if people have the motivation to maintain these behaviors for others' well-being.

# Introduction

During COVID-19 pandemic, especially in the early stages, various preventive behaviors were required because vaccines against COVID-19 had not been developed. Preventive behaviors can be considered a kind of public good that is not sufficiently supplied through market mechanisms where people pursue self-interest [1,2]. Mitigating the pandemic necessitated the collective actions of citizens. However, according to the Peltzman effect, people tend to increase their risky behaviors if safety measures are implemented [3].

Since 2021, vaccines against COVID-19 have been distributed worldwide and have played a vital role in coping with COVID-19. Newly reported cases have decreased in countries where vaccines have been rapidly adopted [4]. People, if they are rational, tend to engage in risky behaviors when security measures are mandated [5]. In terms of economics, this is considered a moral hazard. An empirical question arises as to how the spread of the vaccine influences preventive behaviors [6,7]. As a result of the reduction in the risk of COVID-19 infection, risk-taking behaviors increase, so that preventive behaviors such as staying indoors, wearing masks, and washing hands have changed [8,9]. However, some studies show no clear evidence that vaccinated people have decreased their preventive behaviors in comparison with those not vaccinated [6,10].

The influence of vaccination on preventive behaviors may vary according to the type of behavior [11]. A study found that in China, vaccination lessened the frequency of handwashing but did not change mask-wearing [7]. This study aimed to explore the mechanisms stopping vaccinated people from decreasing preventive behaviors. To this end, we investigated how preventive measures can be pro-socially motivated based on altruism and social solidarity [12].

Using monthly individual-level panel data, we investigated whether individuals' preventive behaviors changed after vaccination. Furthermore, we looked at how the influence of vaccination on preventive behaviors differs according to age and pro-social motivation.

## Methods
### Data collection

We commissioned the research company INTAGE, Inc., to conduct an online survey because of their experience and reliability in academic research. The first wave of queries was conducted March 13–16, 2020, and 4,359 observations were recorded. Participants registered with INTAGE were recruited for this study. The participation rate was 54.7%. The sampling method was designed to collect representatives of the Japanese adult population in terms of educational background, gender, and residential area. For this purpose, INTAGE recruited participants for a survey of pre-registered individuals. However, individuals aged 15 years and below were too young to be registered with INTAGE, and we considered individuals over 80 years of age too old to answer pertinent questions. Inevitably, the sample population was restricted to ages 16–79 years, and participants were randomly selected to fill the pre-specified quotas. INTAGE provided monetary incentives to participants upon study completion.

Internet surveys were conducted repeatedly 26 separate times ("waves") almost every month with identical individuals to construct the panel data. The exceptional period was July-September 2020, when the survey could not be conducted owing to a shortage of research funds. We resumed the surveys after receiving additional funds in October 2020. Vaccination was implemented in April 2021; therefore, the data cover the period before and after implementation of vaccination.

Respondents from the first wave were targeted in subsequent waves to record how some respondents changed their behaviors during the COVID-19 pandemic. In particular, the data allowed us to compare identical persons' preventive behaviors against COVID-19 before and after being vaccinated. During the study period, some identical respondents were dropped from the study sample because some of them stopped taking the surveys, while others did not take the surveys at all. Furthermore, the sample was restricted to those who were completely vaccinated by getting the second shot. In this way, we could compare their behaviors before and after vaccination. Eventually, the number of identical individuals was reduced from 4,359 to 3,019, and the total number of observations used in this study was 70,979.

## Methods

Table 1 presents the key variables used in the estimation. The survey questionnaire contained basic questions about demographics, such as birth year, gender, and educational background. These characteristics were observed at different time points. The surveys were conducted 26 times, between March 2020 and September 2022. During the study period, conditions such as the spread of infection and policies against COVID-19 changed drastically. Table 1 describes the key variables used in the regression estimations. As outcome variables, the respondents were asked questions concerning preventive behaviors, such as:

"Within a week, to what degree have you practiced the following behaviors? Please answer based on a scale of 1 (I have not practiced this behavior at all) to 5 (I have completely practiced this behavior)."

(1) Staying indoors

(2) Wearing a mask

(3) Washing my hands thoroughly

The answers to these questions served as proxies for the following variables for preventive behaviors: staying indoors, frequency of handwashing, and degree of wearing masks. Larger values indicate that respondents are more likely to engage in preventive behaviors. Further, the motivation to get a shot of COVID-19 vaccination is asked in the following question: "Do you get the shot in order to decrease the spread of COVID-19 infection?"

We also asked about the subjective probability of contracting COVID-19 and their perceptions of the severity of COVID-19. We asked whether they received the second shot of the vaccine because the vaccination was effective only after completing the second shot. The latter question was included in the questionnaire from the 12th wave conducted in May 2021 directly after the vaccine was introduced in Japan. The question was included until the 18th wave in November 2021, when most people in the sample completed the second shot. The question was then excluded from the questionnaire, starting with the 19th wave in January 2022.

**Table 1. Definitions of key variables.**

| Variable | Definition |
|---|---|
| | Outcome variables |
| STAYING INDOORS | In the last week, how consistent were you at "not going out of home?" Please choose among 5 choices.<br>1 (not completed at all) to 5 (completely consistent). |
| WEARING MASK | In the last week, how consistent were you at "wearing a mask?" Please choose among 5 choices.<br>1 (not completed at all) to 5 (completely achieved). |
| HANDWASHING | In the last week, how consistent were you at "washing your hands?" Please choose among 5 choices.<br>1 (not completed at all) to 5 (completely achieved). |
| | Confounders (Independent variables) |
| VACCINE | Did you get the second shot?<br>1 (Yes) or 0 (No) |
| PROB_COVID19 | What percentage do you think is the probability of your contracting the novel coronavirus (COVID-19)?<br>0 to 100 (%) |
| SEVERITY COVID19 | How serious do you expect your symptoms to be if you are infected with the novel coronavirus? Choose from 6 choices.<br>1 (very small influence) to 6 (death) |
| AGE_25 | Answer 1 if respondents are age 19–25, otherwise 0 |
| AGE_26_64 | Answer 1 if respondents are age 16–64, otherwise 0 |
| PRO_SOCIAL | In deciding whether you get the shot of COVID-19 vaccine, is it important in preventing the spread of COVID-19?<br>1 (strongly disagree) to 5 (strongly agree).<br>*PRO-SOCIAL* is 1 if respondent chooses 5, otherwise 0. |

We pursued identical respondents from the first-wave survey to the 26th wave for 30 months even though some of the respondents quit the survey. The purpose of this study was to explore how the preventive behaviors of identical persons changed before and after vaccination. Therefore, we limited the sample to those who completed the second shot by the 18th wave and then pursued identical persons until the 26th wave in September 2022. We used panel data containing 3,019 individuals, covering 26 time points from March 2021 to September 2022.

Based on the panel data, we used a fixed-effects (FE) model regression. The FE model is a type of linear regression model widely used in economics. The estimation result using an FE model is equivalent to the results of a linear regression model with dummies for individuals frequently included in each period [13-15]. In this study, 3,019 dummies were included to control for individuals' characteristics that do not change during the period, such as gender, educational background, and childhood experience. Hence, 3,019 confounders were included, reflecting the differences between individuals. Therefore, the estimated results for the time-invariant confounders could not be obtained. Even if various time-variant confounders are included, unobserved individual characteristics cannot be captured. This inevitably results in omitted variable biases [13,15]. For instance, an increasing trend in the number of newly infected persons was observed throughout Japan. This effect is common to all residents in Japan and has changed over time. This can be regarded as a time-fixed effect and can be controlled by including time-period dummies. In this study, 25 time-period dummies were included when one base period was fixed. However, some variables changed not only over time, but also between individuals. Examples are proxy variables for preventive behaviors, which are outcome variables; or number of newly infected persons and deaths due to COVID-19 in residential areas. Furthermore, the timing of obtaining the second vaccine shot is a variable that changed over time and between individuals; therefore, the dummy for vaccination is included in the estimated function as a confounder.

As explained, we controlled not only for unobservable individual fixed effects, but also for unobservable time-fixed effects. This type of FE model is called the two-way error component regression model [15]. This study focused on the correlation between vaccination and preventive behaviors. The statistical software used in this study was the Stata/MP 15.0 multiprocessor from StataCorp, LLC.

The estimated function of an FE model takes the following form:

$$Y_{it} = \alpha_1 \, VACCINE_{it} + \alpha_6 \, PROB \, COVID19_{it} + \alpha_7 \, SEVERITY \, COVID19_{it} + \alpha_8 \, EMERGENCY_{it} + k_t + m_i + u_{it}$$

where Table 1 presents the definitions and basic statistics of these variables. In this formula, $Y_{it}$ represents the outcome variable for individual $i$ and wave $t$, respectively. The time-invariant individual-level fixed effects are represented by $m_i$. Furthermore, $k_t$ represents the effects of different time points, controlled by 25 wave dummies, where the first wave is the reference group. $k_t$ captures various shocks that occurred simultaneously throughout Japan at each time point. Y includes preventive behaviors captured by three proxy variables: *STAYING INDOORS, HANDWASHING*, and *WEARING MASK*. These outcome variables are discrete ordered variables ranging from 1 to 5. Larger values of these variables can be interpreted as indicating that the respondents are more likely to exhibit preventive behavior. In the same specification, we conduct three separate estimations, and the regression parameters are denoted as α. The error term is denoted by $u_{it}$. A simple FE linear regression model is used in this study.

The key confounder is the vaccination dummy; *VACCINE* is 1 if respondents have completed the second shot of the COVID-19 vaccine, otherwise 0. People are obliged to get the second shot within a month of the first shot to make the vaccine effective. Hence, in the sample, there is hardly any time lag between the first and second shots because the survey was conducted every month after the vaccine was approved. There are two age groups: young age (AGE_25) and middle working age (AGE_26_64). The senior group is used as the reference group. PRO_SOCIAL is a dummy variable that captures pro-social motivation to get vaccinated. The mean value of PRO_SOCIAL is 0.86, which shows that 86% of people have the pro-social motivation.

## Results
### Baseline estimations

The coefficient of confounders indicates marginal effects (ME). Table 2 presents the estimation results for the baseline FE model. *VACCINE* shows a positive sign and is statistically significant at the 1% level, with the exception of Column (1), where STAYING INDOORS is the outcome variable. The effects of *VACCINE* are ME 1.036 (95% CI: 0.960-1.111) and ME 0.339 (95% CI: 0.300–0.378) in columns (2) and (3), respectively. Thus, people after vaccination are more likely than before to wear masks by 1.036 points and to wash their hands by 0.339 points, respectively, on a 5-point scale. Before vaccination, the mean values of WEARING MASK and HANDWASHING are 4.39 and 4.15, respectively. Hence, this indicates that the degree of wearing masks increases by 23.6% and washing hands by 8.16% after vaccination, compared with before vaccination. People's behaviors depend on behaviors of others; hence; they follow the social norm [16-19]. Peer pressure is stronger for wearing masks than for washing hands because the surrounding people in a public place can more easily see whether one wears a mask than whether one washes one's hands.

SEVERITY COVID19 shows a significant positive value in all columns, and this means that persons' perception about seriousness of COVID-19 changes the incidence of preventive behaviors.

**Table 2. FE model. Dependent variables are preventive behaviors.**

|  | (1) STAYING INDOORS | (2) WEARING MASK | (3) HANDWASHING |
|---|---|---|---|
| VACCINE | −0.011 (−0.077-0.053) | 1.036*** (0.960-1.111) | 0.339*** (0.300-0.378) |
| PROBABILITY COVID19 | 0.043 (−0.751-0.873) | 0.487 (−0.094-1.070) | 0.560 (0.254-0.867) |
| SEVERITY COVID19 | 0.026*** (0.013-0.039) | 0.029*** (0.017-0.040) | 0.019*** (0.009-0.030) |
| Individual Fixed Effects | Yes | Yes | Yes |
| Time Fixed Effects | Yes | Yes | Yes |
| Controls: New deaths, Newly affected persons, Household income | Yes | Yes | Yes |
| Adjusted $R^2$ | 0.55 | 0.45 | 0.62 |
| Observations | 70,979 | 70,979 | 70,979 |

Note: Numbers within parentheses are 95% CI. For convenience, the coefficient of probability of COVID-19 is multiplied by 1000. The model includes the number of deaths and infected persons in the residential prefectures in the surveys. However, these results have not been reported. "Yes" means that variables are included.

*** $\rho < .01$

## Estimations with cross-terms

The probability and seriousness of contracting COVID-19 differ by age. COVID-19 is more likely to be lethal for adults aged 65 years and older than for younger people [20,21]. Mask-wearing by elderly people is motivated by their self-regarding risk preferences, whereas younger people are motivated by other-regarding concerns [22]. We explore how the effect of COVID-19 vaccination on preventive behaviors differs between age groups. For this purpose, the interaction terms between VACCINE and age groups (AGE_25 and AGE_26_64) are included as key confounders. The reference age group is those over age 65. The results are presented in Table 3. We find a significant negative sign in VACCINE×AGE_25 and VACCINE×AGE_26_64 in column (1), where STAYING INDOORS is the outcome variable. The effects of VACCINE×AGE_25 and VACCINE×AGE_26_64 are ME −0.537 (95% CI: −0.658 - −0.416) and ME −0.295 (95% CI: −0.353 - −0.238). This means that those aged below 25 are less likely to stay home by 0.537 points than those aged over 65, while those aged between 26 and 64 are less likely to stay home by 0.297 points than those aged over 65. However, no differences in the effect of the vaccination are observed when wearing masks and washing hands.

**Table 3. FE model with cross-terms with age cohorts. Dependent variables are preventive behaviors.**

|  | (1) STAYING INDOORS | (2) WEARING MASK | (3) HANDWASHING |
|---|---|---|---|
| VACCINE | 0.206** | 1.058** | 0.319** |
|  | (0.142-0.269) | (0.980-1.137) | (0.281-0.356) |
| VACCINE ×AGE_25 | −0.537*** | −0.066 | −0.056 |
|  | (−0.658 - −0.416) | (−0.160-0.028) | (−0.138-0.024) |
| VACCINE ×AGE_26_64 | −0.295*** | 0.030 | 0.036*** |
|  | (−0.353 - −0.238) | (−0.077-0.015) | (0.004-0.068) |
| Individual Fixed Effects | Yes | Yes | Yes |
| Time Fixed Effects | Yes | Yes | Yes |
| Controls: New deaths, Newly affected persons, Household income | Yes | Yes | Yes |
| PROBABILITY COVID19, SEVERITY COVID19 | Yes | Yes | Yes |
| Adjusted $R^2$ | 0.57 | 0.47 | 0.64 |
| Observations | 70,979 | 70,979 | 70,979 |

Note: Numbers within parentheses are 95% CI. All the models include control variables, which are equivalent to those in Table 2. "Yes" means that variables are included. However, these results have not been reported.

** $p < .01$

*** $p < .05$

We investigated how pro-social motivation affects preventive behaviors. The interaction term between VACCINE and PRO_SOCIAL was included as a key confounder. Table 4 shows the significant positive sign of VACCINE×PRO_SOCIAL, where WEARING MASK and HANDWASHING are the outcome variables. The effects of VACCINE×PRO_SOCIAL are ME 0.110 (95% CI: 0.032 - 0.187) and ME 0.076 (95% CI: 0.014 - 0.137) on WEARING MASK and HANDWASHING, respectively. This suggests that pro-social persons are more likely than non-pro-social persons to wear masks by 0.110 points and to wash their hands by 0.076 points. Effects of VACCINE are ME 0.938 (95% CI: 0.845 -1.031) and ME 0.275 (95% CI: 0.209 - 0.342) on WEARING MASK and HANDWASHING, respectively. This is the effect of vaccination on non-pro-social individuals' preventive behaviors. Considering the results jointly, pro-social persons are 11.8% more likely than non-pro-social persons to wear masks and 27.6% more likely to wash their hands. That is, for pro-social persons, the degree of the effects of vaccination on handwashing is more than twice as great as it is for wearing masks. Wearing masks and washing hands are different in that the benefit of wearing a mask is more likely to depend on the situation. Wearing masks in the open air is only marginally effective in mitigating pandemics [23]. Pro-social vaccinated persons may consider the cost-benefit ratio of preventive behaviors and therefore place more importance on washing hands than on wearing masks.

**Table 4. FE model with cross-term with *PRO_SOCIAL*. Dependent variables are preventive behaviors.**

|  | (1) *STAYING INDOORS* | (2) *WEARING MASK* | (3) *HANDWASHING* |
|---|---|---|---|
| *VACCINE* | −0.051 (−0.154-0.050) | 0.938** (0.845 -1.031) | 0.275** (0.209 - 0.342) |
| *VACCINE ×PRO_SOCIAL* | 0.052 (−0.019 – 0.122) | 0.110** (0.032 - 0.187) | 0.076*** (0.014 - 0.137) |
| *Individual Fixed Effects* | Yes | Yes | Yes |
| *Time Fixed Effects* | Yes | Yes | Yes |
| *Controls: New deaths, New affected persons, Household income* | Yes | Yes | Yes |
| *PROBABILITY COVID19, SEVERITY COVID19* | Yes | Yes | Yes |
| Adjusted $R^2$ | 0.55 | 0.45 | 0.62 |
| Observations | 6,809 | 6,809 | 6,809 |

Note: Numbers within parentheses are 95% CI. All the models include control variables, which are equivalent to those in Table 2. "Yes" means that variables are included. However, these results have not been reported.

** $ρ <.01$

** $ρ < .05$

# Conclusions

In some studies, individuals are unlikely to reduce preventive behaviors even after vaccination [6,10,11]. This is contrary to rational behavior in terms of economics. However, the underlying mechanisms have not yet been investigated. This study contributes to the understanding of the mechanism by considering pro-social motivation. On the one hand, we find that vaccinated people under 65 years of age are less likely to stay indoors than older people. On the other hand, vaccinated individuals are more inclined to wash their hands and wear masks than before being vaccinated. The motivation to be vaccinated, for 86% of respondents, is to mitigate the spread of infection in society. These are considered pro-social people and are more likely than others to wash their hands and wear masks after vaccination. However, their staying-indoors behavior is not different from that of the others.

Staying indoors differs from wearing masks and washing hands when we consider the cost-benefit aspects of preventive behaviors. Staying home leads people to sacrifice their vacation activities in the real world. Using economics terms, the sacrifice is the "opportunity cost" of staying home. They would stay at home if their benefits were greater than their costs. After vaccination, the opportunity cost of staying indoors is higher than its benefit. Accordingly, younger people are more likely to go out.

Both vaccination and preventive behaviors are considered public goods for coping with the pandemic. As a result of vaccination, people tend to go out, which may reduce public goods. To compensate for this, vaccinated pro-social persons are more likely to be motivated to wear masks and wash their hands by considering the benefit to society. Other possible interpretations of the estimation results are related to the cost of getting a vaccination, including the physical and psychological costs of the side effects. From an economic viewpoint, the cost of vaccination can be considered the "sunk cost"—an investment already incurred that cannot be recovered. Due to the sunk cost, vaccinated people continue to invest in public goods by strengthening their mask-wearing and handwashing behaviors rather than their staying-indoors behavior.

Wearing masks in the open air is only marginally effective in mitigating pandemics [23]; thus, this might result in over-investment in public goods. Owing to data limitations, we cannot analyze the situation in which vaccinated people wear masks. It is necessary to determine how and to what extent the preventive behaviors of vaccinated people are effective in mitigating COVID-19.

## List of abbreviations

| | |
|---|---|
| CI | confidence interval |
| COVID-19 | coronavirus-19 |
| FE | fixed effects |
| Inc. | Incorporated |
| LLC | Limited Liability Company |
| ME | marginal effects |